\documentclass{aastex}
\usepackage{emulateapj5,apjfonts,epsfig}

\slugcomment{Accepted for publication in ApJ Letters}
\shorttitle{AXP 4U~0142$+$61}
\shortauthors{JUETT ET AL.}

\begin{document}
\bibliographystyle{apj_noskip}

\title{{\it Chandra\/} High-Resolution Spectrum of the Anomalous X-ray 
	Pulsar 4U~0142$+$61}

\author{Adrienne~M.~Juett,\altaffilmark{1} Herman~L.~Marshall, 
	Deepto~Chakrabarty,\altaffilmark{1,2} \& Norbert~S.~Schulz} 
\affil{\footnotesize Center for Space Research,
	Massachusetts Institute of Technology, Cambridge, MA 02139;\\
	ajuett, hermanm, deepto, nss@space.mit.edu}

\altaffiltext{1}{Also, Department of Physics, Massachusetts Institute of 
	Technology}
\altaffiltext{2}{Alfred P. Sloan Research Fellow}

\begin{abstract}
We report on a 25 ks observation of the 8.7~s anomalous X-ray pulsar 
4U~0142$+$61 with the High Energy Transmission Grating Spectrometer 
(HETGS) on the {\it Chandra X-ray Observatory}.  The continuum spectrum 
is consistent with previous measurements and is well fit by an absorbed 
power-law $+$ blackbody with photon index $\Gamma=3.3\pm0.4$ and 
$kT=0.418\pm0.013$~keV.  No evidence was found for emission or absorption 
lines, with an upper limit of $\approx$50~eV on the equivalent width of 
broad features in the 2.5--13~\AA\/ (0.95--5.0~keV) range and an upper 
limit of $\approx$10~eV on the equivalent width of narrow features in the 
4.1--17.7~\AA\/ (0.7--3.0~keV) range.  If the source is a magnetar, then 
the absence of a proton cyclotron line strongly constrains magnetar 
atmosphere models and hence the magnetic field strength of the neutron 
star.  We find no strong features that are indicative of cyclotron 
absorption for magnetic field strengths of (1.9--9.8)$\times 10^{14}$~G.  
This is still consistent with the dipole field strength of 
$B=1.3\times 10^{14}$~G (at the polar cap) estimated from the pulsar's 
spindown.
\end{abstract}


\section{Introduction}
The anomalous X-ray pulsars (AXPs) have a number of properties that
distinguish them as a class from other pulsars (Mereghetti \& Stella 1995; 
van Paradijs, Taam, \& van den Heuvel 1995).  They have spin periods 
falling in a narrow range (6--12 s), luminosities of order 
10$^{34}$--10$^{35}$ erg s$^{-1}$, and a soft X-ray spectrum well described 
by a blackbody plus a power-law (see Gavriil \& Kaspi 2002 and references 
therein).  They also undergo relatively steady spin-down, have faint or
unidentified optical counterparts, and show no evidence of binary motion.  
Several of the AXPs are associated with supernova remnants, indicating that 
they are relatively young neutron stars. Models to explain these properties 
fall into two general categories: accretion models and magnetar models. 

In the accretion models, the accreting matter does not necessarily
originate from a binary companion.   X-ray timing and optical
counterpart limits allow for only very low-mass companions, although it
is difficult to reconcile such binaries with the supernova remnant
associations because a supernova explosion would likely have 
unbound the binary.   Another accretion model suggests that AXPs are
isolated neutron stars accreting from a fallback disk of material
from the supernova explosion or from the remains of a
Thorne-Zytkow object (e.g., van Paradijs et al. 1995; Ghosh, Angelini, 
\& White 1997).  For all accretion models, there is an equilibrium pulsar 
spin period set by the mass accretion rate and the pulsar magnetic field
strength.  For the observed pulse periods and luminosities, the
derived surface magnetic fields are of order $10^{11}$ G. 

On the other hand, magnetar models have also been proposed, in which
the AXPs are ultramagnetized (B$\sim$10$^{14}$--10$^{15}$ G) neutron
stars whose spin down is due to magnetic dipole radiation.  In these
models, the X-ray luminosity (which far exceeds the spin-down energy)
arises from magnetic field interactions in the neutron star crust
(see, e.g., Thompson \& Duncan 1996).  The spectral and timing
properties of AXPs have been used to constrain models for the
atmospheres of ultramagnetized neutron stars (\"{O}zel, Psaltis, \&
Kaspi 2001).  Other recent work on magnetar atmospheres has shown that
proton cyclotron absorption features should be apparent in the soft
X-ray spectrum of magnetars (Ho \& Lai 2001; Zane et al. 2001).  These lines 
are predicted to be broad ($\Delta E/E\approx1$; Ho \& Lai 2001).  
High-resolution soft X-ray spectroscopy should be able to test the
magnetar model for AXPs. 

The brightest known AXP is 4U~0142$+$61 ($l=129\fdg4$, $b=-0\fdg4$), with 
a spin period of 8.7~s (see, Gavriil \& Kaspi 2002, and references therein).  
Although long ago identified as an unusually soft X-ray source 
(Markert 1976; White \& Marshall 1984; White et al. 1987), confusion with 
the nearby accretion-powered binary pulsar RX~J0146.9$+$6121 contaminated 
the spectrum measured by early X-ray missions.  More recent {\it ASCA\/}
and {\it BeppoSAX\/} spectral studies measured a two component
spectrum consisting of a 0.4~keV blackbody and a $\Gamma$=3.7 photon
power-law (White et al. 1996; Israel et al. 1999; Paul et al. 2000).
Hulleman, van Kerkwijk, \& Kulkarni (2000) identified an
optical counterpart based on the {\it Einstein\/} X-ray position, and 
8.7~s optical pulsations have recently been reported (Kern \& Martin 2001). 

In this {\em Letter}, we discuss our {\it Chandra X-ray 
Observatory\/} High Energy Transmission Grating Spectrometer 
observation of 4U~0142$+$61.  This is the first grating observation 
of an AXP obtained with {\it Chandra}.  We show that the X-ray spectrum 
is consistent with previous lower resolution results and report on 
a search for line features in the spectrum.  We place stringent limits 
on the strength of absorption lines and consequently the allowed magnetic 
field strengths assuming a magnetar model.

\section{Observation and Data Reduction}
We observed 4U~0142$+$61 with {\it Chandra\/} on 2001 May 23 for 25~ks 
using the High Energy Transmission Grating Spectrometer (HETGS) and the 
spectroscopy array of the Advanced CCD Imaging Spectrometer 
(ACIS-S).  For a more detailed description of the instruments, we defer 
to available {\it Chandra\/} X-ray Center (CXC) 
documents\footnote{http://asc.harvard.edu/udocs/docs/docs.html}.  The HETGS 
employs two sets of transmission gratings: the Medium Energy Gratings (MEGs) 
with a range of 2.5--31~\AA\/ (0.4--5.0~keV) and the High Energy Gratings
(HEGs) with a range of 1.2--15~\AA\/ (0.8--10.0~keV).  The HETGS spectra were
imaged by ACIS-S, an array of 6 CCD detectors.  The HETGS/ACIS-S combination 
provides an undispersed (zeroth order) image and dispersed
spectra from the gratings.  The various orders overlap and are
sorted using the intrinsic energy resolution of the ACIS CCDs.  The
first-order HEG (MEG) spectrum has a resolution of $\Delta\lambda=$
0.012~\AA\/ (0.023~\AA).  The ACIS CCDs are normally read out every 3.2~s.  
In order to sample the 8.7-s pulse period of 4U~0142$+$61 more
effectively, we designed the observation to read out only a 512-row
subarray on the ACIS-S detectors, resulting in a frame time of only
1.8~s.  

The ``level 1'' event file was processed using the CIAO v2.1 data
analysis package\footnote{http://asc.harvard.edu/ciao/}.
During our analysis, it was found that the pipeline tool {\tt
acis\_detect\_afterglow} rejects 3--5\% of source photons in grating 
spectra\footnote{http://asc.harvard.edu/ciao/threads/acisdetectafterglow/}.  
Afterglow is the residual charge left from a cosmic-ray event 
which is released over several frames and can cause a line-like feature 
in a grating spectrum; the tool attempts to identify afterglow by flagging 
events that occur at the same chip coordinates in consecutive frames.  
However, for bright sources like 4U~0142$+$61, the tool also rejects a 
small fraction of the valid source events.  Although only a small fraction 
of the total, the rejection of source photons by this tool is systematic 
and non-uniform.  Since order-sorting of grating spectra provides efficient 
rejection of background events, the afterglow detection tool is not 
necessary.  Therefore, we reextracted the event file, retaining those 
events previously tagged by the {\tt acis\_detect\_afterglow} tool in 
order to improve statistics.  No spectral features were found that might 
be attributable to afterglow events. 

The zeroth-order image of 4U~0142$+$61 was affected by photon pileup
(see, e.g., Davis 2001b) and was not used in our spectral  
analysis.  However, we used the zeroth-order image to measure the
source position, with the CIAO tool {\tt celldetect}.  Our best-fit
position was R.A.=$01^{\rm h}\: 46^{\rm m}\: 22\fs44$ and 
Dec.=$61^{\circ}\: 45\arcmin\: 03\farcs3$ (equinox J2000.0), with an
approximate error radius of 0\farcs5.  The optical counterpart of
Hulleman et al. (2000) lies 0\farcs46 from our X-ray position.

The combined MEG and HEG first order count rate was only 4.2~cts~s$^{-1}$.  
We therefore did not expect the dispersed spectrum to be affected by
pileup, although we made several checks nevertheless.  First, the
dispersion distance, $d$, of each event was compared to the energy 
determined by the intrinsic CCD resolution, $E_{\rm ACIS}$.  The grating 
equation predicts $E_{\rm ACIS} = hc/\lambda = hcmR/(Pd)$, where $R$ is 
the Rowland distance, $m$ is the grating order, and $P$ is the grating 
period.  Sources with pileup show a deviation from this relationship in 
lower orders, as low order photons combine to produce higher energy events 
which are then attributed to higher orders.   There were no signs of 
pileup from the observed dispersion-energy relationship.  Also, separate 
model fits of the first and second order spectra were found to be 
consistent both with past (low-resolution) observations and with each 
other, with no signatures of pileup in the residuals.  We conclude that 
pileup is negligible in our dispersed spectra.

We used the standard CIAO tools to create detector response files (ARFs;
see Davis 2001a) for the MEG and HEG $+1$ and $-1$ order spectra. 
These were combined when the $+$/$-$ order spectra were added for the
HEG and MEG separately.  The spectra were binned in two different
ways.  For continuum analysis, we binned the data at 0.08~\AA\/ with a
minimum of 50 counts per bin.  To look for high-resolution spectral
features, the data were binned at 0.015~\AA\/ for the HEG and 0.03~\AA\/ 
for the MEG.  We also created background files for the HEG and
MEG spectra using the standard CIAO tools.  Finally, we used the XSPEC v11
data analysis package (Arnaud 1996) to fit continuum models to the
background-subtracted spectra. 

\centerline{\epsfig{file=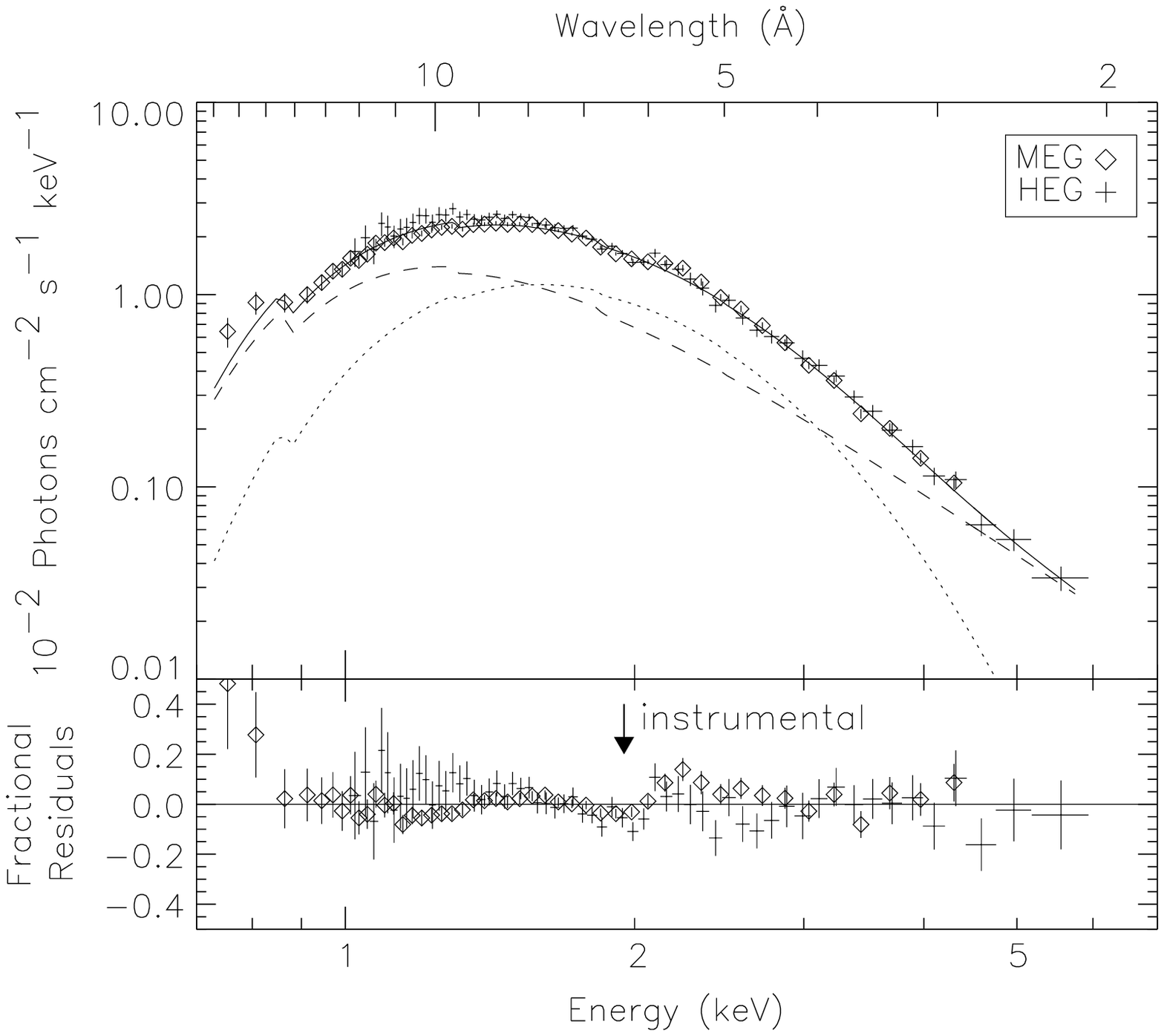,width=0.9\linewidth}}
\figcaption{(upper panel) Unfolded MEG and HEG energy spectra of 4U~0142$+$61 
fit with an absorbed power-law $+$ blackbody model (solid line).  The 
spectra have been coarsely binned for clarity.  The contributions of the 
power-law (dashed line) and blackbody (dotted line) components are 
also shown.  The normalization difference between the MEG and HEG is 
consistent with the uncertainty in the absolute flux calibration.  
(lower panel) Fractional residuals ([data-model]/model) from the absorbed 
power-law $+$ blackbody model fit.  The feature at $\approx$2 keV, marked 
with an arrow, is instrumental.}
\label{fig:1} 
\vspace*{0.1in}

\section{Search for Absorption Features}
We fit a variety of one and two component models to the coarsely binned HEG 
and MEG spectra: power-law, cutoff power-law, thermal bremsstrahlung, 
power-law $+$ blackbody, power-law $+$ disk blackbody, 
power-law $+$ bremsstrahlung, bremsstrahlung $+$ blackbody, and cutoff 
power-law $+$ blackbody.  All models included absorption by neutral 
interstellar gas with cosmic abundances.  The {\it Chandra\/} best fit 
parameters of the most acceptable models are given in Table 1.  

\begin{figure*}[t]
\centerline{\epsfig{file=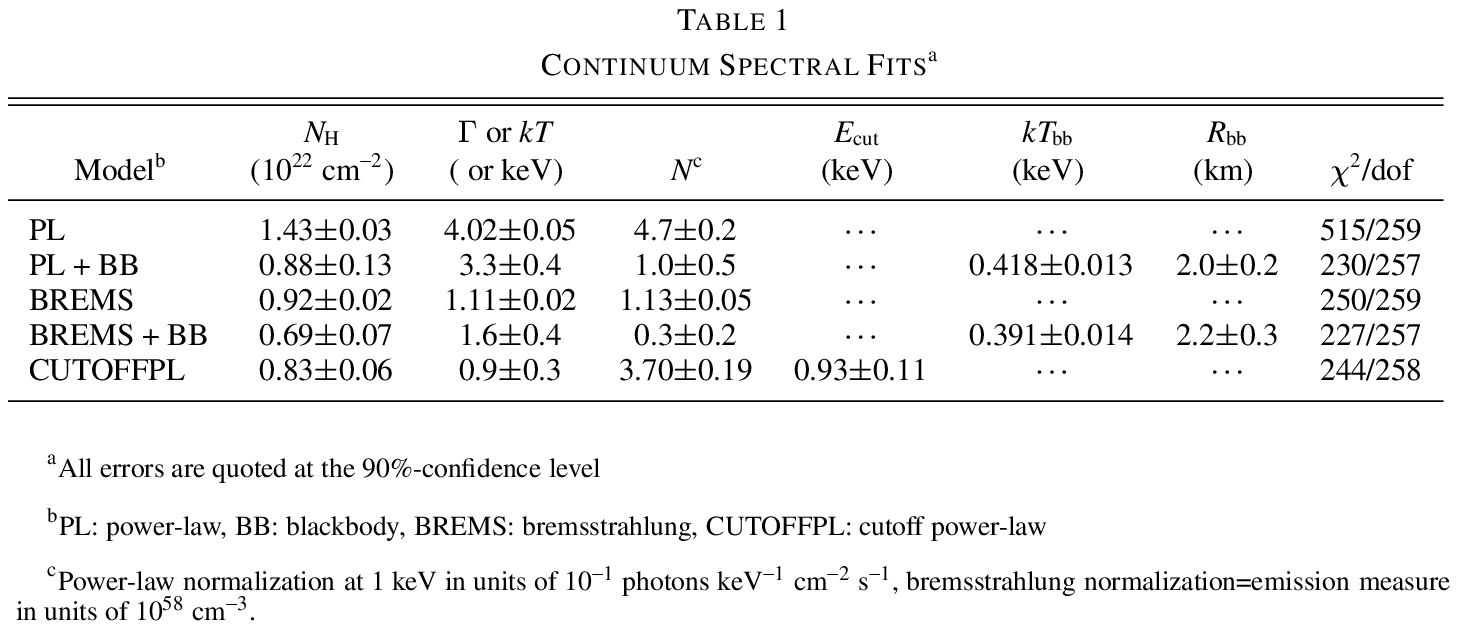}}
\end{figure*}

The cutoff power-law $+$ blackbody best fit parameters were inconsistent 
and gave a cutoff energy far above the instrument bandpass, suggesting 
that a standard power-law model was sufficient.  The one component models 
had larger reduced $\chi^2$ values than the two component models and 
showed residuals that suggested the need for a second component.  Using 
an {\it F}-test (see, e.g., Bevington \& Robinson 1992), we determined that 
the addition of a blackbody to the power-law and bremsstrahlung models is 
highly significant.  Therefore, we rule out the single component models: 
power-law, cutoff power-law, and thermal bremsstrahlung.  The two component 
bremsstrahlung models all gave large emission measures 
$\approx 10^{58}$ $d_{\rm kpc}^{2}$ cm$^{-3}$ which would require an 
emission radius orders of magnitude larger than a neutron star for an 
optically thin plasma (White et al. 1996), and were rejected on this basis.  
Traditionally, the source has been fit with an absorbed power-law 
$+$ blackbody model and our analysis favors this model as well.  The 
results of the {\it Chandra\/} fit, summarized in Table 1 and shown in 
Figure 1, are consistent with previous observations of 4U~0142$+$61.  
When the same model is applied to the high-resolution binned data set, 
the same parameter values are found to within the uncertainties.

\centerline{\epsfig{file=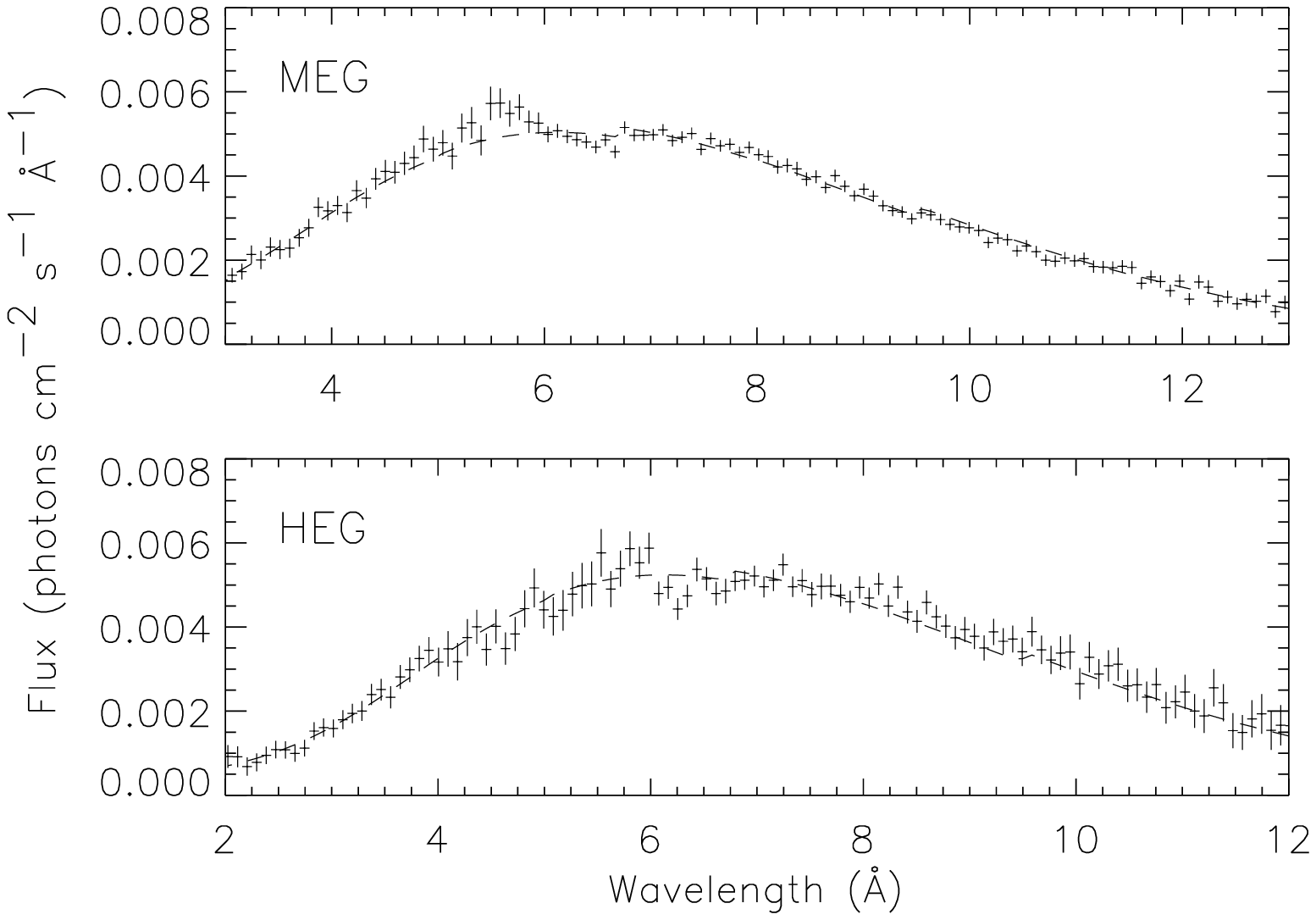,width=0.9\linewidth}}
\figcaption{MEG and HEG spectra of 4U~0142$+$61 at 6$\times$ and 12$\times$ 
instrument resolution, respectively.  The best fit absorbed 
power-law $+$ blackbody model is shown as a dashed line.  The misfit of 
the continuum between 5.5--7.0 \AA\/ is not consistent in the MEG and HEG 
spectra, and is attributed to instrumental features around known absorption 
edges.  The spectra show no deep features that might result from proton 
cyclotron resonance absorption in a magnetic field in the range of 
(1.9--9.8)$\times 10^{14}$ G.}
\label{fig:2} 
\vspace*{0.1in}
\vspace*{\fill}

With this continuum model, we then looked for absorption 
and emission features in the high-resolution spectrum.  We searched for 
both narrow and broad line-like features.  An initial inspection of the 
residuals of the power-law $+$ blackbody model fit seemed to show a 
feature at $\approx$6~\AA\/ (2 keV; see Figure 1).  This feature is however 
highly suspect, since it is located between two known instrumental edges 
(Si-K at 6.75~\AA\/ and Ir-M at 5.98~\AA).  It does not appear in the 
MEG $+$1 spectrum (which is on a back-illuminated chip) but only in the 
MEG $-$1 and HEG $+$1/$-$1 spectra (on front-illuminated chips).  
Furthermore, the dip occurs at a slightly different energy and with a 
different depth in the MEG and HEG spectra (see Figure 2).  For these 
reasons, we conclude that the feature is instrumental.

To search for broad features, we used the high-resolution HETGS spectra and 
fit Gaussian line models to the fractional residuals ([data-model]/model) 
of the continuum fit.  The central energies and widths of the Gaussian 
components were fixed, while the normalizations were fitted.  To look for 
features that had been predicted in the magnetar models (Ho \& Lai 2001; 
Zane et al. 2001), the sigma of the Gaussian was chosen to vary with 
energy ($\sigma =0.1\times E$).  The Gaussian model was fit to the data 
centering at every wavelength point.  We were able to determine the best 
fit amplitude and standard deviation of this result as well as the 
significance of each feature.  The 4$\sigma$ upper limits on the 
equivalent widths (EWs) of any features are shown in Figure 3.  There were no 
features with a significance greater than 4$\sigma$, which we believe is 
a reasonable lower limit.  In Figure 3, we have also included lines which 
mark the equivalent width limits for EW=$0.1E$ and EW=$0.5E$ for comparison.  

\centerline{\epsfig{file=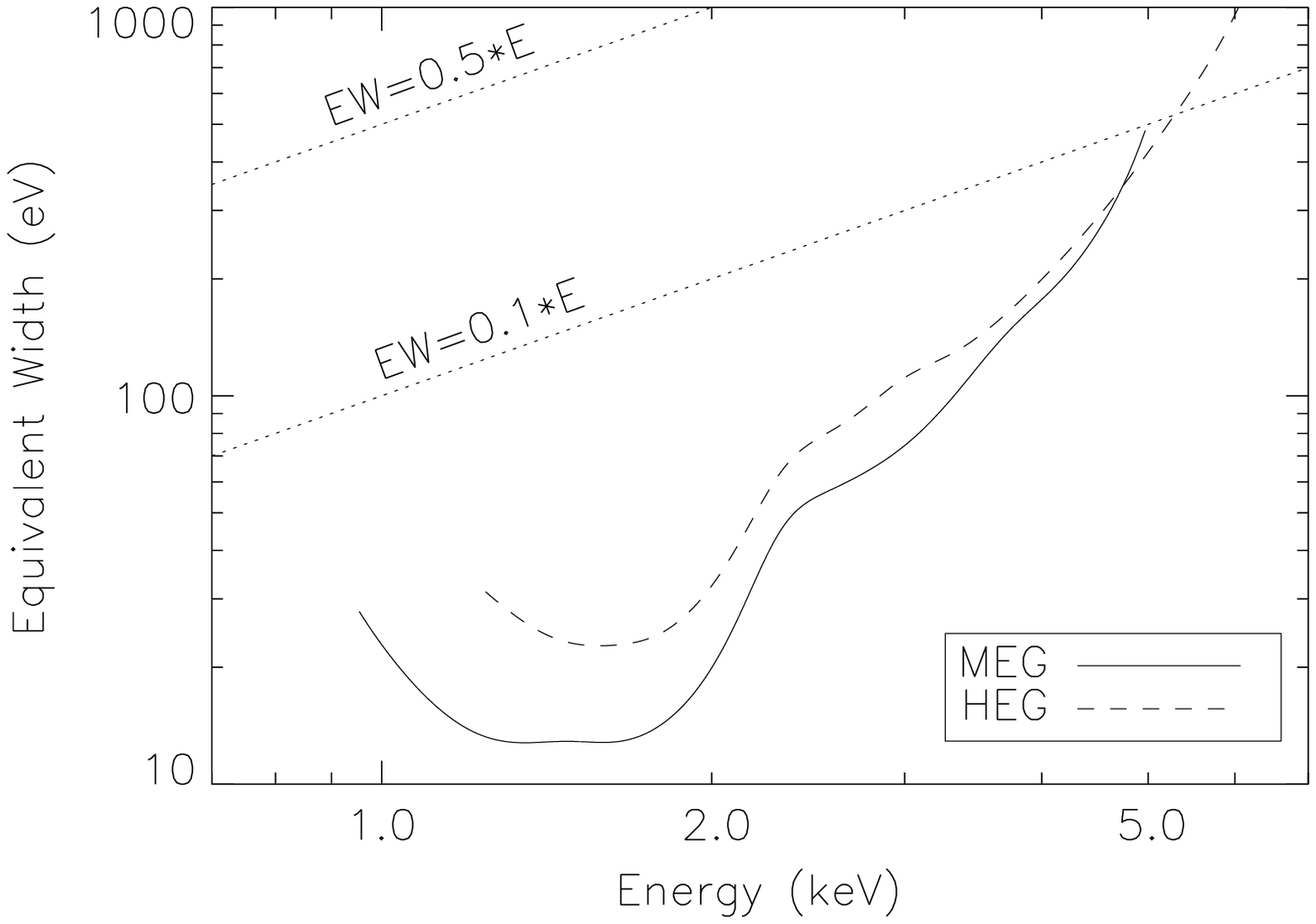,width=0.9\linewidth}}
\figcaption{Upper bounds on the equivalent widths of lines with intrinsic 
width $\sigma=0.1E$ (where $E$ is the line energy).  Magnetar atmosphere 
models predict that the equivalent width of a proton cyclotron absorption 
line is of order 0.70--0.75$E$ (Zane et al. 2001), although more recent work 
predicts equivalent widths at least an order of magnitude lower (Ho \& Lai 
2002).  For comparison, we have included lines which define EW=$0.5E$ and 
EW=$0.1E$.  We conclude that the spectra show no strong absorption feature 
attributable to the proton cyclotron resonance in a magnetic field in the 
range of (1.9--9.8)$\times 10^{14}$ G.}
\label{fig:3} 
\vspace*{0.1in}
\vspace*{\fill}

Similarly, no narrow line features were found in the high-resolution 
spectra.  We can place a 3$\sigma$ upper limit of 10 eV on any emission 
line between 4.1--17.7 \AA\/ (0.7--3.0~keV) whose width (FWHM) is comparable 
to the instrument resolution (see Figure 4).  The computation uses the model 
fit to the time-averaged HEG and MEG data.  We assume that candidate features 
are only 2 bins wide.  One may derive limits on broader features until 
the scale of the feature becomes comparable to that of the instrument 
calibration uncertainties.  Limits on absorption lines are identical when 
there are $\gtrsim$25 counts per bin but are systematically larger for 
$\lesssim$25 counts per bin.  

\centerline{\epsfig{file=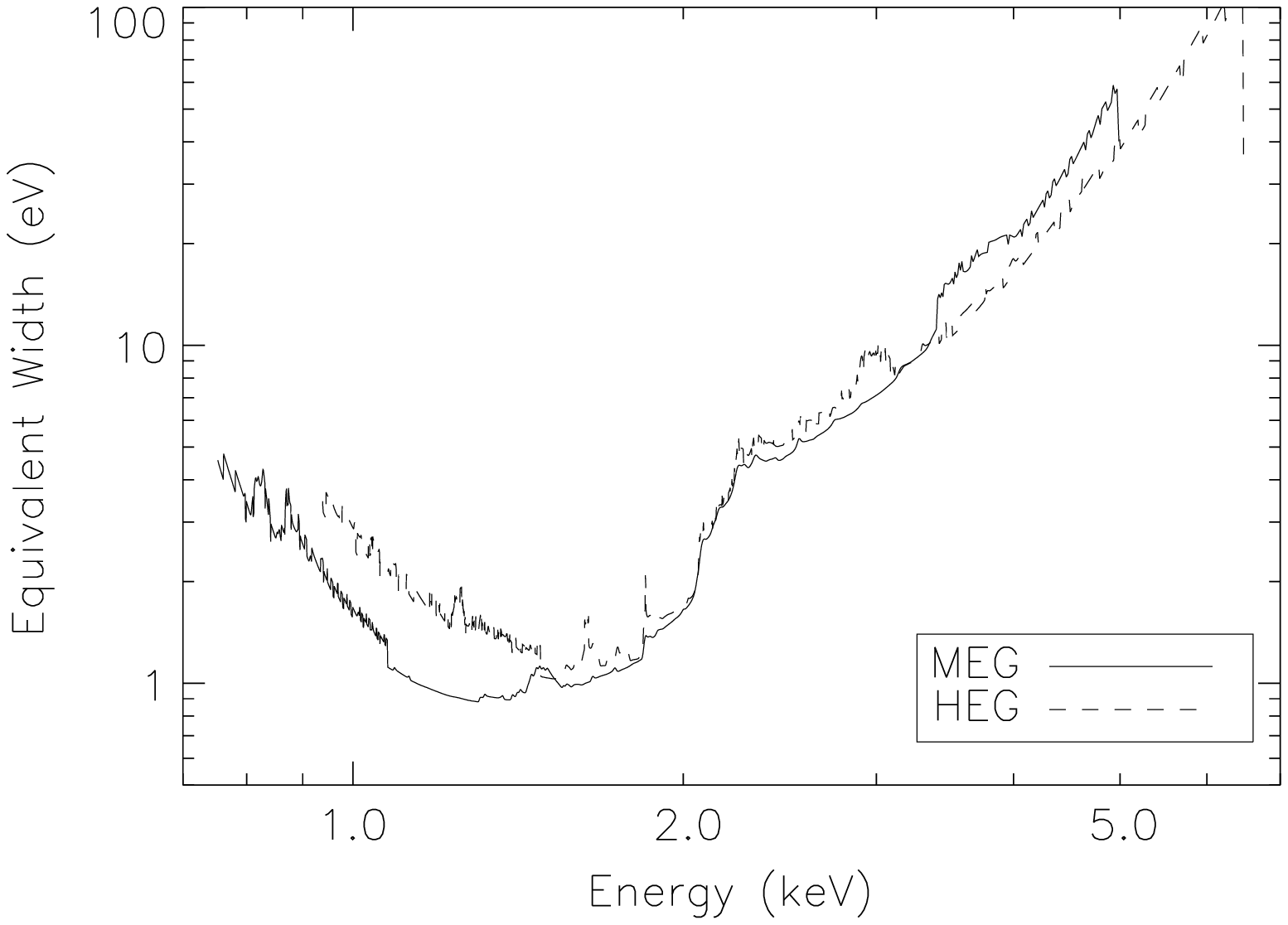,width=0.9\linewidth}}
\figcaption{Upper bounds on the equivalent widths of narrow lines with 
FWHMs comparable to the instrument resolution, 0.023 \AA\/ (0.012 \AA), for 
MEG (HEG).  We assumed that candidate features are only 2 bins wide.  One 
may derive limits on broad features using these curves until the scale of 
the feature becomes comparable to that of the instrument calibration 
uncertainties.  Limits on absorption lines are identical when there are 
$\gtrsim$25 counts per bin but are systematically larger for $\lesssim$25 
counts per bin.}
\label{fig:4} 
\vspace*{0.1in}

\section{Discussion}
The {\it Chandra\/} spectrum of 4U~0142$+$61 is well fit by an 
absorbed power-law $+$ blackbody, which is consistent with previous 
observations of the source (White et al. 1996; Israel et al. 1999; 
Paul et al. 2000).  The small radius of the blackbody model might be 
explained by atmospheric effects which give an overestimation of the 
temperature and thus an underestimation of the radius when fit by a 
standard blackbody (see, e.g., Perna et al. 2001, and references therein).  
The best fit hydrogen column density of 
(0.88$\pm$0.13)$\times 10^{21}$ cm$^{-2}$ is consistent with estimates 
derived from radio and optical studies (Dickey \& Lockman 1990; 
Hulleman et al. 2000).  Distance estimates from extinction measurements 
for 4U~0142$+$61 range from 1--5 kpc due to the line of sight passing 
through the edge of a local ($<1$ kpc) molecular cloud.  When including 
effects of the molecular cloud, a lower limit of 1 kpc is determined, 
whereas when neglecting the molecular cloud a lower limit of 2.7 kpc is 
obtained (\"{O}zel et al. 2001).  Normalizing the distance to 1 kpc, we 
find a luminosity (0.5--10~keV) of 
5.3$\times 10^{34}$ $d_{\rm kpc}^{2}$~ergs~s$^{-1}$ which is 
consistent with past measurements from {\it ASCA\/} and {\it BeppoSAX\/}.

No significant discrete features were found in the high-resolution 
spectrum. We can place a limit on the equivalent width of any narrow feature 
of $\approx$10~eV in the range 4.1--17.7~\AA\/ (0.7--3.0~keV) and 
$\approx$50~eV for broad features in the range 2.5--13~\AA\/ (0.95--5.0~keV).  
The limits on the equivalent width of any absorption features place 
strong constraints on the neutron star atmosphere models.  Atmosphere 
models of ultramagnetized neutron stars predict a broad proton cyclotron 
absorption feature at an energy $E_{B}=0.63y_{g}(B/10^{14}\:{\rm G})$~keV, 
where $y_{g}=(1-2GM/c^{2}R)^{1/2}$ is the gravitational redshift factor 
and $B$ is the surface magnetic field strength (Ho \& Lai 2001; Zane et al. 
2001).  If we assume that magnetic dipole radiation drives the spindown, 
the magnetic field strength found using {\it P} and {\it \.{P}} 
($B^{2}\propto P \dot{P}$) is 1.3$\times 10^{14}$ G (Gavriil \& Kaspi 
2002).  A dipole field produces a cyclotron absorption feature consistent 
with a surface magnetic field strength 20--30\% lower (Zane et al. 2001).  
For 4U~0142$+$61, the dipole field of 1.3$\times 10^{14}$ G would give a 
cyclotron absorption feature at $\approx$0.5 keV, which is well outside 
of the range accessible to our observation.  

Initial theoretical work predicted a proton cyclotron absorption feature 
of equivalent width 0.70--0.75$E_{B}$ (Ho \& Lai 2001; Zane et al. 2001), 
which is much greater than the equivalent width limits in the 0.95--5.0~keV 
range.  If we take $y_{g}\sim0.8$, the above energy range corresponds to 
magnetic field strengths of (1.9--9.8)$\times 10^{14}$ G.  More recent 
work has shown that vacuum polarization effects strongly suppress the 
cyclotron absorption feature, giving equivalent widths an order of 
magnitude lower (Ho \& Lai 2002).  It should also be noted that the 
theoretical calculations of the equivalent width of the cyclotron feature 
were done for a local patch of the neutron star surface.  A phase averaged 
spectrum, like the one we are using, would include contributions from 
various magnetic field strengths, directions and effective temperatures, 
which would further suppress the cyclotron feature.  The combination of 
these effects may reduce the predicted equivalent widths to below our 
measured values for much of the spectral range we studied. 

\acknowledgements{We thank Claude R. Canizares, the HETGS principal 
investigator, for allocating part of his {\it Chandra\/} guaranteed time 
program to this observation.  We also thank Wynn C. G. Ho, Dong Lai, 
Feryal \"{O}zel, Dimitrios Psaltis, and Peter Woods for useful discussions.
This research was supported in part by 
contracts SAO SV1-61010 and NAS8-38249 as well as NASA grant NAG5-9184.}


\begin{thebibliography}{0}
\parskip=0pt
\parsep=0pt
\itemsep=0pt

\bibitem[Arnaud (1996)]{a96}
Arnaud, K. A. 1996, in Astronomical Data Analysis Software and Systems V, 
ed. G. Jacoby \& J. Barnes (San Francisco: ASP Conf. Ser. 101), 17

\bibitem[Bevington \& Robinson (1992)]{br92}
Bevington, P. R., \& Robinson, D. K. 1992, Data Reduction and Error 
Analysis for the Physical Sciences, 2nd ed. (New York: McGraw-Hill)

\bibitem[Davis (2001a)]{d01a}
Davis, J.~E. 2001a, ApJ, 548, 1010

\bibitem[Davis (2001b)]{d01b}
---------. 2001b, ApJ, 562, 575

\bibitem[Dickey \& Lockman (1990)]{dl90}
Dickey, J. M., \& Lockman, F. J. 1990, Ann. Rev. Ast. Astr., 28, 215

\bibitem[Gavriil \& Kaspi (2002)]{gk02}
Gavriil, F. P., \& Kaspi, V. M. 2002, ApJ, in press (astro-ph/0107422)

\bibitem[Ghosh, Angelini, \& White (1997)]{gaw97}
Ghosh, P., Angelini, L., \& White, N. E. 1997, ApJ, 478, 713

\bibitem[Ho \& Lai (2001)]{hl01}
Ho, W. C. G., \& Lai, D. 2001, MNRAS, 327, 1081

\bibitem[Ho \& Lai (2002)]{hl02}
Ho, W. C. G., \& Lai, D. 2001, MNRAS, submitted (astro-ph/0201380)

\bibitem[Hulleman, van Kerkwijk, \& Kulkarni (2000)]{hvk00}
Hulleman, F., van Kerkwijk, M. H., \& Kulkarni, S. R. 2000, Nature, 408, 689

\bibitem[Israel et al. (1999)]{ioa+99}
Israel, G. L. et al. 1999, A\&A, 346, 929

\bibitem[Kern \& Martin (2001)]{km01}
Kern, B., \& Martin, C. 2001, IAU Circ., No. 7769

\bibitem[Markert (1976)]{m76}
Markert, T. H. 1976, PhD thesis, Massachusetts Institute of Technology

\bibitem[Mereghetti \& Stella (1995)]{ms95}
Mereghetti, S., \& Stella, L. 1995, ApJ, 442, L17

\bibitem[\"{O}zel (2001)]{o01}
\"{O}zel, F. 2001, ApJ, 563, 276

\bibitem[\"{O}zel, Psaltis, \& Kaspi (2001)]{opk01}
\"{O}zel, F., Psaltis, D., \& Kaspi, V. M. 2001, ApJ, 563, 255

\bibitem[Paul et al. (2000)]{pkd+00}
Paul, B., Kawasaki, M., Dotani, T., \& Nagase, F. 2000, ApJ, 537, 319

\bibitem[Perna et al. (2001)]{phh+01}
Perna, R., Heyl, J. S., Hernquist, L. E., Juett, A. M., \& Chakrabarty, D. 
2001, ApJ, 557, 18

\bibitem[Thompson \& Duncan (1996)]{td96}
Thompson, C., \& Duncan, R. C. 1996, ApJ, 473, 322

\bibitem[Thorne \& Zytkow (1977)]{tz77}
Thorne, K. S., \& Zytkow, A. N. 1977, ApJ, 212, 832

\bibitem[van Paradijs, Taam, \& van den Heuvel (1995)]{vtv95}
van Paradijs, J., Taam, R. E., \& van den Heuvel, E. P. J. 1995, A\&A, 299, L41

\bibitem[White et al. (1996)]{wae+96}
White, N. E., Angelini, L., Ebisawa, K., Tanaka, Y., \& Ghosh, P. 1996, 
ApJ, 463, L83

\bibitem[White \& Marshall (1984)]{wm84}
White, N. E., \& Marshall, F. E. 1984, ApJ, 281, 354

\bibitem[White et al. (1987)]{wmg+87}
White, N. E., et al. 1987, MNRAS, 226, 645

\bibitem[Zane et al. (2001)]{zts+01}
Zane, S., Turolla, R., Stella, L., \& Treves, A. 2001, ApJ, 560, 384

\end{thebibliography}
\end{document}